\documentclass[aps,prd,
nofootinbib,
twocolumn,
floatfix
]{revtex4-1}

\usepackage{epsfig}
\usepackage{amsmath}
\usepackage{amssymb}
\usepackage{graphicx}

\newcommand{\aap}{{A\&A}}%

\begin{document}

\title{Offset of the dark matter cusp and the interpretation of the
  130\,GeV line as a dark matter signal}

\author{Dmitry Gorbunov} 
\affiliation{Institute for Nuclear Research of the Russian Academy of
  Sciences,\\60-th October Anniversary pr. 7a, 117312 Moscow, Russia} 
\affiliation{Moscow Institute of Physics and Technology, \\Institutsky
  per. 9, Dolgoprudny, 141700, Russia} 
\author{Peter Tinyakov}
\affiliation{Service de Physique Th\'{e}orique, Universit\'{e} Libre
  de Bruxelles (ULB),\\CP225 Boulevard du Triomphe, B-1050 Bruxelles,
  Belgium}

\begin{abstract} 
We show that the cusp in the dark matter (DM) distribution required to explain
the recently found excess in the gamma-ray spectrum at energies $\sim 130$~GeV
in terms of the DM annihilations cannot survive the tidal forces if it is
offset by $\sim 1.5^\circ$ from the Galactic center as suggested by
observations.
\end{abstract}

\maketitle

\section{Introduction}
Recently, a line-like feature in the gamma-ray spectrum observed by the Fermi
LAT experiment in the direction of the Galactic center has been found.  It has
been suggested that it can be interpreted in terms of the dark matter (DM)
annihilation signal \cite{Bringmann:2012vr,Weniger:2012tx}.

More refined analysis has confirmed the significance of the feature in the
spectrum \cite{Tempel:2012ey,Su:2012ft}. The best fit to the DM annihilation
line was obtained with the modified Navarro-Frenk-White (NFW) DM profile with
the inner slope of $\alpha=-1.2$ (as may have resulted, 
e.g., from the adiabatic contraction). It also revealed the offset of the signal
with respect to the Galactic center by about $1.5^\circ$, which corresponds to
the projected distance of $\sim 200$\,pc.  The existence of the offset, if
confirmed, raises a question about the viability of the DM interpretation of
the observed gamma-ray excess.

In order to address this question, numerical simulations of the Galactic
center including the effect of the bar were used in Ref.\,\cite{Kuhlen:2012qw}.
It was found that an offset of a few hundred parsec between the GC and the
maximum of the DM distribution could in principle exist (see also
Ref.~\cite{deBlok:2009sp} and 
references therein). However, the DM
distribution in this case is not cuspy but cored
and has an overdensity with respect to
the central region of only about $10-20$\%. In addition, it was found
that the core density is formed by the gravitationally unbound DM particles.

Here we address the same question from a different perspective. 
 Rather than trying to constrain possible mechanisms by which an offset cusp
could be produced (one may think, e.g., of a merger event in the past), we consider how
long such a cusp would survive. 
Making use of
the analytical estimates we show that only a small central region of the DM
cusp can survive the tidal forces produced by the baryons which dominate the
gravitational potential near the Galactic center. The surviving part of the
cusp is too small to explain the observed feature in the $\gamma$-ray
spectrum.

\section{Baryonic and DM profiles in the Galactic center} 

The baryon distribution in the GC is known from the 2~$\mu$m light
distribution (see, e.g., 
Refs.~\cite{1989ApJ...338..824M,1990ApJ...359..112S,2002A&A...384..112L})
and confirmed by the study of the kinematic properties of the OH/IR stars
\cite{1992A&A...259..118L}.

For our purposes the exact behavior of the baryon density is not necessary,
and a crude approximation in the inner $\sim 200$~pc is sufficient. From
Fig.~10 of Ref.~\cite{1992A&A...259..118L} we adopt the following approximation
for the baryonic mass $M_B(r)$ enclosed within the radius $r$,
\begin{equation}
M_B(r) = 6\times 10^8 \,M_\odot \left({r\over 100~{\rm pc}}\right)^{1.25}\;,
\label{eq:MofR}
\end{equation} 
where $M_\odot$ is the solar mass. 
>From this equation, the baryonic density $\rho_B(r)$ is 
\begin{align*}
\rho_B(r) &= 60 M_\odot \,{\rm pc}^{-3} \left({r\over 100~{\rm
    pc}}\right)^{-1.75}\\
&= 2.3\times 10^3\,  {\rm GeV\,cm}^{-3}\left({r\over 100~{\rm pc}}\right)^{-1.75}.
\end{align*}
This relation is consistent, within the errors, with the one derived
in Refs.~\cite{1968ApJ...151..145B,1990ApJ...359..112S}. It is clear from
eq.~(\ref{eq:MofR}) that at distances $\gtrsim 100$\,pc from the Galactic
center the effect of the central black hole is subdominant, and we ignore it
in what follows. 

Now we turn to the DM distribution. In Ref.~\cite{Su:2012ft} several
DM profiles were shown to fit the gamma-ray data, namely the Einasto
profile \cite{Einasto} and modified NFW profiles of the form 
\begin{equation}
\rho(r) = {\rho_s\over (r/r_s)^\alpha (1+r/r_s)^{3-\alpha}},
\label{eq:NFW}
\end{equation}
where $r_s=20$\,kpc and the slope $\alpha$ ranges form $\alpha =1$ to
$\alpha=1.3$, with the best fit value $\alpha=1.2$. The normalization factor
$\rho_s = 0.27$\,GeV/cm$^3$ is fixed by setting the DM density around the
Earth to $0.4$\,GeV/cm$^3$. To illustrate our point it is sufficient to
consider the simpler and more cuspy NFW profile (\ref{eq:NFW}). The less cuspy
Einasto profile is subject to even stronger tidal effects.

As a first step, let us find the size of the region which is responsible for
the gamma-ray signal assuming the latter is produced by the DM
annihilations. The observed signal corresponds to the luminosity
\cite{Su:2012ft}
\[
L_0 = (1.7\pm0.4)\times 10^{36} {\rm photons/s}.
\]
On the other hand, the luminosity of a spherical region of the size
$r$ centered on the DM distribution can be written as follows, 
\begin{equation}
L(r) = {4\pi r_s^3\over m_{\rm DM}^2}\, \langle\sigma v\rangle\, \rho_s^2\, 
I(r/r_s)\;,
\label{eq:luminosity}
\end{equation}
where $\langle\sigma v\rangle \simeq 2\times 10^{-27}$\,cm$^3$/s is the
velocity-averaged DM annihilation cross section \cite{Weniger:2012tx}, $m_{\rm
  DM}$ is the DM mass ($m_{\rm DM}\simeq 130$\,GeV
\cite{Weniger:2012tx,Su:2012ft} in the case of annihilation
into $\gamma\gamma$ and $140$-$150$\,GeV
\cite{Bringmann:2012vr,Tempel:2012ey} in the case of internal bremsstrahlung)
and
\begin{equation}
I(r/r_s) = \int_0^{r/r_s} {x^2 \, dx \over x^{2\alpha} 
(1+x)^{6-2\alpha}}\;.
\label{eq:IofR}
\end{equation}
Equating this to the observed luminosity gives the 
following equation for the size $r$ of the emission region, 
\[
I(r/r_s) = {L_0\, m_{\rm DM}^2\over 4\pi\, \langle\sigma v\rangle
  \,\rho_s^2 \,
  r_s^3 } = 7.7\times 10^{-2}\;. 
\]
This equation is easily solved by noting that the solution corresponds
to small values of $r/r_s$ for which the integral in
eq.\,\eqref{eq:IofR} can be simplified, 
\[
I(r/r_s) \simeq {1\over 3-2\alpha} (r/r_s)^{3-2\alpha}\;.
\]
For the best fit case $\alpha=1.2$ this gives 
\begin{equation}
r \simeq 0.006\, r_s = 120\,{\rm pc},
\label{eq:rL}
\end{equation}
where we have used $m_{\rm DM}=140$\,GeV. Note that when obtaining this
estimate we have assumed (cf. eq.~(\ref{eq:luminosity})) that all of the DM
mass is converted into photons, as in the case of the annihilation into
$\gamma\gamma$. If the efficiency were lower, as it would be in the case of
the annihilation into a single-photon final state
\cite{Tempel:2012ey,Bringmann:2012vr}, the size of the contributing region
would be even larger.

It is instructive to calculate the total amount of DM contained within
a given distance from the cusp. Making use of the relation
$r/r_s\ll 1$ one finds 
\[
M_{\rm DM}(r) = 4\pi \rho_s \, r_s^3 \, {1\over 3-\alpha} \,
(r/r_s)^{3-\alpha}\,, 
\]
where $r$ is the distance form the center of the cusp. 
This implies for $\alpha=1.2$ 
\begin{equation}
M_{\rm DM}(r) = 2.9\times 10^7 M_\odot \left({r\over 100\,{\rm
    pc}}\right)^{1.8}.
\label{eq:mu(r)}
\end{equation}
The latter value has to be compared to eq.\,(\ref{eq:MofR}). Clearly, at
distances $\sim 200$~pc from the Galactic center the baryons give a dominant
contribution to the total mass. Thus, they dominate the gravitational
potential except in the vicinity of the cusp.

\section{Tidal stripping of the offset DM cusp}

The DM cusp that is offset with respect to the baryonic distribution is
subject to tidal forces. To estimate the importance of these forces one may
compare the difference of the gravitational pull of baryons at different parts
of the DM distribution and the gravitational force from the DM itself. For the
cusp to survive the gravitational force from baryons has to be smaller than
the force from DM. This leads to the condition
\begin{equation}
{G M_{\rm DM}(d) \over d^2} \gtrsim { GM_B(r) \over r^3}\, d\;,
\label{eq:stability}
\end{equation}
where $r\sim200$\,pc is the offset distance and it was assumed that $d\ll r$.
Making use of eqs.~(\ref{eq:MofR}) and (\ref{eq:mu(r)}) one finds for
$\alpha=1.2$, $d \lesssim 22\,{\rm pc}$. Thus, the tidal radius (the maximum
radius where DM particles survive the stripping by tidal forces) is much
smaller than the offset distance, which justifies the approximation used.

In fact, the value obtained from eq.~(\ref{eq:stability}) is an
overestimate. More accurately, the tidal radius $d$ can be calculated by
making use of the formalism developed in Ref.~\cite{Read:2005zm}. In
principle, one should distinguish tidal radii corresponding to prograde,
radial and retrograde DM orbits. However, in the long time limit (that is, at
times much longer than the period of the orbital motion $\sim 10^7$~yr) these
converge to the smallest of the three. Making use of eq.~(21) of
Ref.~\cite{Read:2005zm}, one finds at $\alpha=1.2$
\begin{equation}
d\simeq 5\,{\rm pc}\;.
\label{eq:tidalR}
\end{equation}
At smaller values of $\alpha$ the DM cusp is weaker and the tidal radius is
slightly smaller, while at larger values of $\alpha$ it is slightly larger,
always being of the same order as given by eq.~(\ref{eq:tidalR}). Thus, only a
very small central part of the cusp can survive the tidal disruption. 
This part is much smaller than the region responsible for the annihilation
signal.

\section{Conclusions} 

>From the mismatch between eqs.\,(\ref{eq:rL}) and (\ref{eq:tidalR}) it is
clear that the part of the cusp which can survive the tidal stripping is
insufficient to explain the DM signal. First, its angular size is about $2'$
that is smaller than the Fermi-LAT point-spread
function\,\cite{Ackermann:2012qk}.  Such a source would appear as point-like,
which is not compatible with the morphology of the observed excess.

Second, according to eq.~(\ref{eq:luminosity}), the annihilation signal from
such a small  region \eqref{eq:tidalR}
would be reduced by a factor of about $\sim 7$ if the
annihilation cross section of $\langle\sigma v\rangle \simeq 2\times
10^{-27}$~cm$^3$/s is assumed. To make the signal from the surviving part of
the cusp compatible with observations, one would have to increase the cross
section by the same factor, which would be in contradiction with the limits
from Fermi LAT \cite{Ackermann:2012qk} (note that the latter are integral 
limits which are insensitive to the contribution from the small region
around the Galactic center).

\acknowledgments

We are indebted to A.\,Frolov, M.\,Gustafsson, M.~Libanov, B.\,Torsten and
C.~Weniger for helpful discussions. We would also like to thank the
hospitality of the JINR-ISU Baikal Summer School where this work was
initiated. The work of D.G. is supported in part by the grant of the President
of the Russian Federation NS-5590.2012.2, by RFBR grant 11-02-01528-a, by MSE
under contract \#8412 and by SCOPES program.  D.G. acknowledges the
hospitality of the Service de Physique Th\'eorique of ULB where this work was
completed.  The work of P.T. is supported in part by the IISN, the ULB-ARC and
the Belgian Science Policy (IAP VII/37).


\begin{thebibliography}{99}
\bibitem{Bringmann:2012vr} 
  T.~Bringmann, X.~Huang, A.~Ibarra, S.~Vogl and C.~Weniger,
  JCAP {\bf 1207}, 054 (2012)
  [arXiv:1203.1312 [hep-ph]].

\bibitem{Weniger:2012tx} 
  C.~Weniger,
  JCAP {\bf 1208}, 007 (2012)
  [arXiv:1204.2797 [hep-ph]].

\bibitem{Tempel:2012ey} 
  E.~Tempel, A.~Hektor and M.~Raidal,
  JCAP {\bf 1209}, 032 (2012)
  [Addendum-ibid.\  {\bf 1211}, A01 (2012)]
  [arXiv:1205.1045 [hep-ph]].

\bibitem{Su:2012ft} 
  M.~Su and D.~P.~Finkbeiner,
  arXiv:1206.1616 [astro-ph.HE].

\bibitem{Kuhlen:2012qw} 
  M.~Kuhlen, J.~Guedes, A.~Pillepich, P.~Madau and L.~Mayer,
  arXiv:1208.4844 [astro-ph.GA].

\bibitem{deBlok:2009sp} 
  W.~J.~G.~de Blok,
  Adv.\ Astron.\  {\bf 2010}, 789293 (2010)
  [arXiv:0910.3538 [astro-ph.CO]].

\bibitem{1989ApJ...338..824M}
\bibinfo{author}{M.~T. {McGinn}}, \bibinfo{author}{K.~{Sellgren}},
  \bibinfo{author}{E.~E. {Becklin}}, and \bibinfo{author}{D.~N.~B. {Hall}},
  \bibinfo{journal}{\apj} \bibinfo{volume}{\textbf{338}}, \bibinfo{pages}{824}
  (\bibinfo{date}{Mar. 1989}).

\bibitem{1990ApJ...359..112S}
\bibinfo{author}{K.~{Sellgren}}, \bibinfo{author}{M.~T. {McGinn}},
  \bibinfo{author}{E.~E. {Becklin}}, and \bibinfo{author}{D.~N. {Hall}},
  \bibinfo{journal}{\apj} \bibinfo{volume}{\textbf{359}}, \bibinfo{pages}{112}
  (\bibinfo{date}{Aug. 1990}).

\bibitem{2002A&A...384..112L}
\bibinfo{author}{R.~{Launhardt}}, \bibinfo{author}{R.~{Zylka}}, and
  \bibinfo{author}{P.~G.~{Mezger}}, \bibinfo{journal}{\aap}
  \bibinfo{volume}{\textbf{384}}, \bibinfo{pages}{112} (\bibinfo{date}{Mar. 2002}).

\bibitem{1992A&A...259..118L}
\bibinfo{author}{M.~{Lindqvist}}, \bibinfo{author}{H.~J. {Habing}}, and
  \bibinfo{author}{A.~{Winnberg}}, \bibinfo{journal}{\aap}
  \bibinfo{volume}{\textbf{259}}, \bibinfo{pages}{118} (\bibinfo{date}{Jun.
  1992}).

\bibitem{1968ApJ...151..145B}
  \bibinfo{author}{E.~E. {Becklin}}, and \bibinfo{author}{G. {Neugebauer}},
  \bibinfo{journal}{\apj} \bibinfo{volume}{\textbf{151}}, \bibinfo{pages}{145}
  (\bibinfo{date}{1968}).

\bibitem{Einasto} J. Einasto (1965), 
Kinematics and dynamics of stellar systems, 
Trudy Inst. Astrofiz. Alma-Ata 51, 87

\bibitem{Read:2005zm} 
  J.~I.~Read, M.~I.~Wilkinson, N.~W.~Evans, G.~Gilmore and J.~T.~Kleyna,
  Mon.\ Not.\ Roy.\ Astron.\ Soc.\  {\bf 366}, 429 (2006)
  [astro-ph/0506687].

\bibitem{Ackermann:2012qk} 
  M.~Ackermann {\it et al.}  [LAT Collaboration],
  Phys.\ Rev.\ D {\bf 86}, 022002 (2012)
  [arXiv:1205.2739 [astro-ph.HE]].

\end{thebibliography}

\end{document}